\documentclass[twocolumn,english,aps,floatfix,showpacs,prb]{revtex4-1}
\usepackage{amsmath}
\usepackage{amssymb}
\usepackage{graphicx}
\usepackage{color}
\usepackage{dcolumn}
\usepackage{bm}
\usepackage{ulem}
\usepackage{babel}
\usepackage{babel}
\usepackage{babel}
\usepackage{babel}
\usepackage{babel}
\usepackage{babel}
\usepackage{babel}

\makeatletter

\@ifundefined{textcolor}{}{
 \definecolor{BLACK}{gray}{0}
 \definecolor{WHITE}{gray}{1}
 \definecolor{RED}{rgb}{1,0,0}
 \definecolor{GREEN}{rgb}{0,1,0}
 \definecolor{BLUE}{rgb}{0,0,1}
 \definecolor{CYAN}{cmyk}{1,0,0,0}
 \definecolor{MAGENTA}{cmyk}{0,1,0,0}
 \definecolor{YELLOW}{cmyk}{0,0,1,0}
 }
\@ifundefined{textcolor}{}{
 \definecolor{BLACK}{gray}{0}
 \definecolor{WHITE}{gray}{1}
 \definecolor{RED}{rgb}{1,0,0}
 \definecolor{GREEN}{rgb}{0,1,0}
 \definecolor{BLUE}{rgb}{0,0,1}
 \definecolor{CYAN}{cmyk}{1,0,0,0}
 \definecolor{MAGENTA}{cmyk}{0,1,0,0}
 \definecolor{YELLOW}{cmyk}{0,0,1,0}
 }
\@ifundefined{textcolor}{}{
 \definecolor{BLACK}{gray}{0}
 \definecolor{WHITE}{gray}{1}
 \definecolor{RED}{rgb}{1,0,0}
 \definecolor{GREEN}{rgb}{0,1,0}
 \definecolor{BLUE}{rgb}{0,0,1}
 \definecolor{CYAN}{cmyk}{1,0,0,0}
 \definecolor{MAGENTA}{cmyk}{0,1,0,0}
 \definecolor{YELLOW}{cmyk}{0,0,1,0}
 }
\@ifundefined{textcolor}{}{
 \definecolor{BLACK}{gray}{0}
 \definecolor{WHITE}{gray}{1}
 \definecolor{RED}{rgb}{1,0,0}
 \definecolor{GREEN}{rgb}{0,1,0}
 \definecolor{BLUE}{rgb}{0,0,1}
 \definecolor{CYAN}{cmyk}{1,0,0,0}
 \definecolor{MAGENTA}{cmyk}{0,1,0,0}
 \definecolor{YELLOW}{cmyk}{0,0,1,0}
 }
\makeatother

\begin{document}

\title{Tensor network algorithm \\
by coarse-graining tensor renormalization on finite periodic lattices}
\author{Hui-Hai Zhao$^{1,2}$ }
\email{zhaohuihai@solis.t.u-tokyo.ac.jp}
\author{Zhi-Yuan Xie$^{3,4}$}
\author{Tao Xiang$^{3,5}$}
\email{txiang@iphy.ac.cn}
\author{Masatoshi Imada$^{1}$}
\email{imada@ap.t.u-tokyo.ac.jp}
\affiliation{$^1$Department of Applied Physics, University of Tokyo, Hongo, Bunkyo-ku,
Tokyo 113-8656, Japan}
\affiliation{$^2$Institute for Solid State Physics, University of Tokyo, Kashiwanoha,
Kashiwa, Chiba 277-8581, Japan}
\affiliation{$^3$Institute of Physics, Chinese Academy of Sciences, P.O. Box 603,
Beijing 100190, China}
\affiliation{$^4$Department of Physics, Renmin University of China, Beijing 100872, China}
\affiliation{$^5$Collaborative Innovation Center of Quantum Matter, Beijing 100190, China}
\date{\today }

\begin{abstract}
We develop coarse-graining tensor renormalization group algorithms to
compute physical properties of two-dimensional lattice models on finite
periodic lattices. Two different coarse-graining strategies, one based on
the tensor renormalization group and the other based on the higher-order
tensor renormalization group, are introduced. In order to optimize the
tensor-network model globally, a sweeping scheme is proposed to account for
the renormalization effect from the environment tensors under the framework
of second renormalization group. We demonstrate the algorithms by the
classical Ising model on the square lattice and the Kitaev model on the
honeycomb lattice, and show that the finite-size algorithms achieve
substantially more accurate results than the corresponding infinite-size
ones.
\end{abstract}

\maketitle

\section{Introduction\label{sec:Introduction}}

Tensor network algorithms, which have been widely developed recently \cite%
{Niggemann1997,PTP2001Nishino,PEPS2004,MERA,TRG2007,PRL2007Vidal,SimpleUpdate2008,TEFR,SRG2009,SingleLayer2011,HOTRG2012, PESS2014,TNR}%
, provide efficient and promising numerical tools for studying two and
higher dimensional classical statistical and quantum lattice models. For a
classical statistical model\cite
{Book1982Baxter,TRG2007,SRGfull2010,PRE2008Berker} or a lattice gauge model%
\cite{PRE2014Yujifeng,PRD2014Alan}, the partition function can be expressed
as a tensor network state, and physical quantities can be obtained by
contracting this tensor network state. In the quantum case, the tensor
network state provides a faithful representation of the ground state wave
function. For example, the ground state of Affleck-Kennedy-Lieb-Tasaki model%
\cite{AKLT}, which is the so-called valence-bond-solid state, can be
rigorously represented as a tensor network state. 

A tensor-network wave
function was first used in the numerical calculation of the Heisenberg model
on the honeycomb lattice by Niggemann \textit{et al} in 1997\cite%
{Niggemann1997}. Later on, Sierra \textit{et al} proposed a more general
ansatz on the tensor network states\cite{Sierra1998}. By utilizing tensor
network state as a variational wave function, Nishino \textit{et al}
proposed a method to calculate three-dimensional Ising model\cite%
{PTP2001Nishino}. In 2004, Verstraete and Cirac argued that a natural
generalization of the matrix product state to two and higher dimensions is a
projected entangled pair state (PEPS)\cite{PEPS2004}, and showed that this
kind of states satisfies automatically the area law of entanglement entropy.
As a generalization of the PEPS, the projected entangled simplex state
(PESS) \cite{PESS2014} was proposed by Xie \textit{et al} to account for the
entanglement effect of all particles in a simplex. This simplex entanglement
effect is important in the study of frustrated quantum lattice systems. By
introducing disentanglers to the tensor network to reduce the short range
entanglement, the multi-scale entanglement renormalization ansatz\cite{MERA}
can provide logarithmic correction to the area law of entanglement entropy%
\cite{RMP2010Eisert} in one-dimensional critical systems. Tensor network
algorithms have not only shown prominent success in the study of both
classical statistical and quantum lattice models, but also drawn intense
interests in the study of quantum field theory\cite%
{PRL2010Cirac,PRB2014Zueco}, interacting fermions\cite%
{PRA2010Cirac,arXiv2010Wen,PRB2015Kao,arXiv2015Eisert}, and disordered
systems\cite{PRB2014Zhou}.

During the past years, several algorithms have been developed to contract
two-dimensional infinite lattice tensor network states. These algorithms can
be divided into two categories. The first is to represent a tensor-network
state as a product of transfer matrix along a spatial direction, and reduce
the problem of evaluating the trace of the tensor-network state to a problem
of finding the dominant eigenvalues of the transfer matrix. To diagonalize
the transfer matrix, a number of approaches can be used. One is the transfer-matrix renormalization group\cite{JPSJ1995Nishino,TMRG1996,PRB1997Xiang} as
well as its generalization such as the corner-transfer-matrix
renormalization group\cite%
{MP1968Baxter,NishinoCTMRG1996,OrusCTMRG2009,PRB2010Corboz}, the other is
the infinite time-evolving block decimation based on the evolution of matrix
product state\cite{PRL2007Vidal,QIC2007Verstraete,canonical2008}.

The second is to contract the tensor network state by taking a coarse
graining tensor renormalization. This includes the method of tensor
renormalization group (TRG)\cite{TRG2007,TERG2008} method proposed by Levin
and Nave based on the singular value decomposition (SVD) of local tensors.
The TRG is a local optimization approach, which optimizes the truncation
error only locally. A global optimization approach of TRG, developed by
Xie \textit{et al}. \cite{SRG2009,SRGfull2010}, was proposed to take into account for
the second renormalization effect of the environment tensors. This method is
called the second renormalization group (SRG) and can significantly reduce
the truncation error in comparison with the TRG. In 2012, a different
coarse-graining scheme based on the higher-order singular value
decomposition (HOSVD), which is now referred as HOTRG\cite{HOTRG2012,CPL2014Wang}, was
proposed. Similarly to the TRG, the HOTRG is a local optimization method.
The corresponding global optimization method is called higher-order second
renormalization group (HOSRG)\cite{HOTRG2012}.

The above coarse graining TRG methods allow us to evaluate thermodynamic
quantities directly in the thermodynamic limit, without invoking finite-size
scaling. However, the truncation error is accumulated at each
coarse-graining step. The accuracy of the method is reduced with the
increase of the lattice size, especially in a system close to a critical
point where the correlation length diverges. The error can be reduced by
increasing the bond dimension $D$ of local tensors retained in the
truncation. Furthermore, a finite $D$ scaling analysis can be performed to
extrapolate to the infinite $D$ limit.\cite{PRB2014Nishino}

The PEPS algorithm was first introduced by Verstraete and Cirac to study finite lattice systems in 2004\cite{PEPS2004}, 
and was further improved by a number of groups
\cite{PEPS2007,PEPS2008,finitePEPS2014,
Recycle2014,PEPSfiniteT2015,arXiv2015Orus}. 
It works more efficiently in a system with open boundary conditions than in a system with periodic boundary conditions, due to the heavy computational cost in the evaluation of a double-layer tensor network state in a periodic system.
The coarse graining TRG or SRG can be applied to a finite periodic system\cite{TNMC2011,PRB2015Kao}.
But a systematical study on this kind of methods is still not available.

In this paper, we explore the finite-lattice SRG algorithm in a periodic boundary system. We propose a sweeping scheme to evaluate the environment tensor using the SRG. It is found that the sweeping can greatly improve the accuracy of the SRG, especially in the case the truncation error is large. As for the Kitaev model considered in this work,
we find that the SRG with sweeping can significantly reduce the truncation error comparing to that of the TRG.
It suggests that the SRG can be used to determine very accurately the renormalization effect of the environment in a finite lattice system by sweeping. This, together with the finite-size scaling analysis of the finite-lattice results, gives a powerful numerical tool for studying lattice models in two or higher dimensions.

This paper is arranged as follows.
In order to make the paper as self-contained as possible, we first briefly review the TRG, SRG and other related methods introduced in Refs. \onlinecite{PEPS2004,TRG2007,SRG2009,SRGfull2010,HOTRG2012}.
In Sec. \ref{sec:ConstructTN}, the construction of tensor network from classical statistical models and quantum lattice models are described. In Sec. \ref{sec:TRG}, the TRG and HOTRG methods are briefly reviewed.
In Sec. \ref{sec:SRG}, as an example,
the implementation of the SRG on the honeycomb lattice, and the HOSRG on the square lattice are illustrated. In Sec. \ref{sec:finiteSRG}, the application of SRG and HOSRG on finite lattices are explained. In Sec. \ref{sec:benchmark}, we benchmark our algorithms for the classical Ising model on the square lattice and the Kitaev model on the honeycomb lattice. Finally, we summarize in Sec. \ref{sec:summary}.

\section{Tensor network state\label{sec:ConstructTN}}

\begin{figure}[tbp]
\includegraphics[width=7cm]{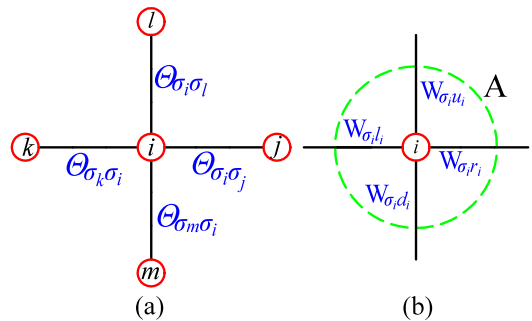}
\caption{(color online) (a) Pictorial representation of the partition
function defined by Eq.~(\protect\ref{eq:IsingPartition}), where $\protect%
\Theta_{\protect\sigma_{i}\protect\sigma_{j}}$, $\protect\Theta_{\protect%
\sigma_{k}\protect\sigma_{i}}$,$\dots$ are the Boltzmann weights associated
with neighboring sites. (b) Every Boltzmann weight is decomposed as in Eq.~(%
\protect\ref{eq:decompBoltz}), and a site tensor $A$ is constructed from 4 $W
$ matrices associated with the same site as in Eq.~(\protect\ref%
{eq:siteTensor}). }
\label{fig:IsingTN}
\end{figure}

As elaborated in Ref. \onlinecite{SRGfull2010}, the partition function of
any classical statistical lattice model with short-range interactions can be
represented as a tensor network state. A simple example is the two-dimensional Ising model, defined by the Hamiltonian
\begin{equation}
H=-\sum_{\left\langle i,j\right\rangle }\sigma_{i}\sigma_{j},
\end{equation}
where $\sigma_{i}=\pm1$ is the Ising variable at site $i$. The partition function of this model is given by
\begin{equation}
Z=\sum_{\left\{ \sigma\right\} }\prod_{\left\langle i,j\right\rangle
}\exp\left(\beta \sigma_{i}\sigma_{j} \right),  \label{eq:IsingPartition}
\end{equation}
where $\sum$ is to sum over all spin configurations $\left\{ \sigma\right\}$, and $\beta$ is the inversed temperature. The Boltzmann weight associated with site $i$ and $j$ (Fig. \ref{fig:IsingTN}(a)) can be decomposed as\cite{HOTRG2012}
\begin{equation}
\Theta_{\sigma_{i}\sigma_{j}} = e^{\beta \sigma_{i}\sigma_{j}} =
\sum_{u_i} W_{\sigma_{i} u_i}W_{\sigma_{j} u_{i}}.  \label{eq:decompBoltz}
\end{equation}
Here, $W$ is $2\times2$ matrix, defined by
\begin{equation}
W=
\begin{pmatrix}
\sqrt{\cosh\beta}, & \sqrt{\sinh \beta} \\
\sqrt{\cosh \beta} & -\sqrt{\sinh \beta}%
\end{pmatrix}%
,
\end{equation}
where $u_{i}$ is the index representing the bond basis state between sites $i$ and $j$.
On square lattice, A local tensor at site $i$ can then be constructed as
\begin{equation}
A_{u_{i}d_{i}l_{i}r_{i}}^{i}=\sum_{\sigma_{i}}
W_{\sigma_{i}u_{i}}W_{\sigma_{i}d_{i}}W_{\sigma_{i}l_{i}}W_{\sigma_{i}r_{i}},
\label{eq:siteTensor}
\end{equation}
where $u_i, d_i, l_i$ and $r_i$ represents the bonds between the $i$ site
and its up, down, left and right nearest neighbor sites, respectively.
A graphical representation of this equation is shown in Fig. \ref{fig:IsingTN}(b).

%==========================================================
%==========================================================

By substituting the above equations to Eq. (\ref{eq:IsingPartition}), it is straightforward to show that the partition function of the Ising model on a square lattice can be expressed as a product of local tensors(Fig. \ref{fig:IsingTN})\cite{SRGfull2010}
\begin{equation}
Z=\mathrm{Tr}\prod_{i}A_{u_{i}d_{i}l_{i}r_{i}}^{i},  \label{eq:TNpartition}
\end{equation}
where $i$ runs over all the lattice sites and the trace is to sum over all bond indices.

For a quantum lattice model, the ground state wave function can also be expressed as a tensor network state. Both PEPS and PESS belong to this type of wave functions. Let us take a square-lattice PEPS
 functions\cite{PEPS2004} as an example:
\begin{equation}
\left|\Psi\right\rangle =\sum_{\left\{ m\right\} } \mathrm{Tr}%
\prod_{i}a_{u_{i}d_{i}l_{i}r_{i}}^{i}\left[m_{i}\right]
\left|m_{1}m_{2}\ldots m_{N}\right\rangle ,  \label{eq:PEPS}
\end{equation}
where $i$ runs over all the lattice sites. The trace is to sum over all bond indices. The dimension of each bond index is assumed to be $D$. The difference
between the PEPS wave function in Eq.~(\ref{eq:PEPS}) and the tensor network
representation of the partition function in Eq.~(\ref{eq:TNpartition}) is
that Eq.~(\ref{eq:PEPS}) contains extra indices $m_{1}m_{2}\ldots m_{N}$ which represent physical degrees of freedom on all lattice sites.

The PEPS is a variational wave function for the ground state. Unlike the tensor network states for classical statistical models, the tensor elements of PEPS are unknown. The PEPS can be determined by variationally minimizing the ground state energy. But the computational cost for doing this is very high. A more efficient approach for determining the PEPS is to take an imaginary time evolution by applying the projection operator $\exp (-\tau H)$ successively to a random PEPS, where $\tau$ is a small parameter.
For a Hamiltonian, which contains only nearest neighbor interactions
\begin{equation}
H=\sum_{\left\langle i,j\right\rangle }h_{ij},
\end{equation}
the imaginary time evolution generally starts by taking the Suzuki-Trotter decomposition to divide approximately the evolution operator into a sequence of local two-site operators
\begin{equation}
e^{-\tau H}= \prod_{\left\langle i,j\right\rangle
}e^{-\tau h_{ij}} + O\left(\tau^{2}\right).
\end{equation}
The imaginary time evolution is then implemented by applying
these two-site projection operators to the PEPS successively.

\begin{figure}[tbp]
\includegraphics[width=8.5cm]{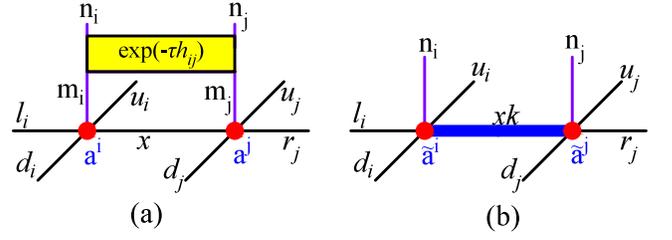}
\caption{(color online) (a) Pictorial representation of applying the
operator $\exp\left(-\protect\tau h_{ij}\right)$ on the associated
neighboring site tensors $a^{i}$ and $a^{j}$. (b) Site tensors $\tilde{a}^i$
and $\tilde{a}^j$ after the evolution by Eq. (\protect\ref{eq:ai_tilde}) and
(\protect\ref{eq:aj_tilde}), where the dimension of thicker blue bond is
increased. }
\label{fig:imagTimeEvo}
\end{figure}

To do the projection, let us first do the following singular value decomposition for the two-site projection operator
\begin{equation}
\left\langle m_{i}m_{j}\right| e^{-\tau
h_{ij}} \left|n_{i}n_{j}\right\rangle
=\sum_{k}U_{m_{i}n_{i},k}S_{k}V_{m_{j}n_{j},k},
\end{equation}
where $U$ and $V$ are unitary matrices of dimension $d^2$ with $d$ the dimension of local physical basis space. $S$ is a semi-positive diagonal matrix, and its diagonal elements are the singular values of the projection operator.
By applying the projection operator $\exp\left(-\tau h_{ij}\right)$ to the PEPS, the local tensors at sites $i$ and $j$ change to
\begin{eqnarray}
\tilde{a}_{udl\left(xk\right)}^{i}\left[m\right]
& = & \sum_{n}a_{udlx}^{i}\left[n\right] U_{mn,k}%
\sqrt{S_{k}}  ,\label{eq:ai_tilde}
\\
\tilde{a}_{ud\left(xk\right)r}^{j}\left[m\right]
& = & \sum_{n}a_{udxr}^{j}\left[n\right] V_{mn,k}%
\sqrt{S_{k}}.  \label{eq:aj_tilde}
\end{eqnarray}
The bond dimension on the link connecting $i$ and $j$ are enlarged by a factor of $d^2$. A graphical representation of these equations is shown in Fig. \ref{fig:imagTimeEvo}.

%===================
%===================

To continue the imaginary time evolution, the bond dimension on the link between $i$ and $j$ needs to be truncated.
The truncated local tensors, $a^i$ and $a^j$, can be determined by minimizing the following difference between the tensors before and after the truncation \cite{PEPS2004}
\begin{equation}
  f (a) = \left|\left|\Psi\right\rangle -\left| \tilde{\Psi}\right\rangle \right|^{2} ,
  \label{Eq:min}
\end{equation}
where $\left|\tilde{\Psi}\right\rangle $ and $\left| \Psi\right\rangle $ are the PEPS wave functions before and after truncations, respectively.
$f(a)$ is a quadratic function of $a^{i}$ and $a^{j}$.
It can be generally expressed as\cite{PEPS2007}
\begin{equation}
f\left(a\right)=a^{\dagger}\mathcal{N}a-2a^{\dagger}b+\mathrm{const},
\label{eq:quadratic}
\end{equation}
where $a$ represents $a^i$ or $a^j$. $\mathcal{N}$ is obtained from the norm $\left\langle \Psi|\Psi\right\rangle $ by removing $a^{\dagger}$ in $\left\langle \Psi\right|$ and $a$ in $\left\langle \Psi\right|$, and $b$ is obtained from
$\left\langle \Psi|\tilde{\Psi}\right\rangle $ by removing $a^{\dagger}$ in $%
\left\langle \Psi\right|$.

To explicitly show how $\mathcal{N}$ and $b$ are derived from the PEPS wave
function, let us first define the following double tensors
\begin{eqnarray}
A_{\mu\mu^{\prime}}& = &\sum_{m}a_{udlr}^{*}\left[m\right] a_{u^{\prime}d^{%
\prime}l^{\prime}r^{\prime}}\left[m\right],  \label{eq:doubleTensor}
\\
\tilde{A}_{\mu\mu^{\prime}}& = &\sum_{m}\tilde{a}_{udlr}^{*}\left[m\right] a_{u^{\prime}d^{%
\prime}l^{\prime}r^{\prime}}\left[m\right],
\end{eqnarray}
where $\mu$ and $\mu^{\prime}$ are the simplified notations of bond indices $%
udlr$ and $u^{\prime}d^{\prime}l^{\prime}r^{\prime}$, respectively. Then, $%
\mathcal{N}$ and $b$ tensor networks corresponding to site $i$ can be
expressed as
\begin{eqnarray}
\mathcal{N}_{\mu_{i}m_{i},\mu_{i}^{\prime}m_{i}^{\prime}}^{i} &= & \delta_{m_{i},m_{i}^{\prime}} \mathrm{%
Tr}\prod_{q\neq i}A_{\mu_{q}\mu_{q}^{\prime}}^{q}
,
\\
b_{\mu_{i}m_{i}}^{i}& =& \mathrm{Tr} \tilde{a}_{\mu_{i}^{\prime}}^{i}%
\left[m_{i}\right] \tilde{A}_{\mu_{j}\mu_{j}^{\prime}}^{j} \prod_{q\neq i,j}A_{\mu_{q}\mu_{q}^{\prime}}^{q} ,
\end{eqnarray}
where Tr is to contract all connecting bond indices.

$\left\langle \Psi|\Psi\right\rangle $ and $\left\langle \Psi|\tilde{\Psi}%
\right\rangle $ are called closed tensor networks since they are composed of
closed bonds, and the contraction of the closed tensor network comes out to
be a scalar. $\mathcal{N}$ and $b$ are called open tensor networks since
they are composed of open bonds in addition to the closed bonds, and the
contraction of a open tensor network comes out to be a tensor, which is
usually called the environment tensor. In Eq.~(\ref{eq:quadratic}), $%
\mathcal{N}$ is the environment tensor of $a^{\dagger}$ and $a$, and $b$ is
the environment tensor of $a^{\dagger}$.

By minimizing the functional $f(a)$, one can update the local tensors. This procedure is called full-update\cite{PEPS2004,PEPS2007}. An entanglement mean-field approach can also be used to update the local tensors without contracting all tensors\cite{SimpleUpdate2008}. This can avoid the contraction of the full double tensor network, which requires demanding computation time.
Nevertheless, this kind of contraction is still needed in the evaluation of expectation value
\begin{equation}
\left\langle \mathcal{O}\right\rangle = \frac{\langle
\Psi\left|\mathcal{O}\right|\Psi \rangle}{\left\langle
\Psi|\Psi\right\rangle} .
\end{equation}

\section{Coarse graining tensor renormalization\label{sec:TRG}}

The previous section describes the construction of the tensor network.
Then, the following task is to perform the network contraction,
which can be achieved by coarse-graining tensor renormalization.
The general idea of these methods is to
reduce number of tensors by a coarse-graining renormalization procedure
iteratively until the tensor network becomes small enough to be exactly
contracted.
Levin and Nave\cite{TRG2007} proposed the TRG to study two-dimensional classical models based on the SVD of local tensors.
Later, Xie et al\cite{HOTRG2012} proposed the HOTRG to study two and three dimensional lattice models based on the HOSVD of local tensors.
Both of these two methods optimizes the truncation error only locally, which do not consider the contribution from environment lattice sites.
We briefly reiterate the procedures of TRG and
HOTRG in this section.

\subsection{Tensor renormalization group}

We here take a two-sublattice honeycomb-lattice
tensor network as an example to describe the TRG method. The TRG method on the square lattice is described in Appendix \ref{apdx:squareTRG}.

%=========================
\begin{figure}[tbp]
\includegraphics[width=8cm]{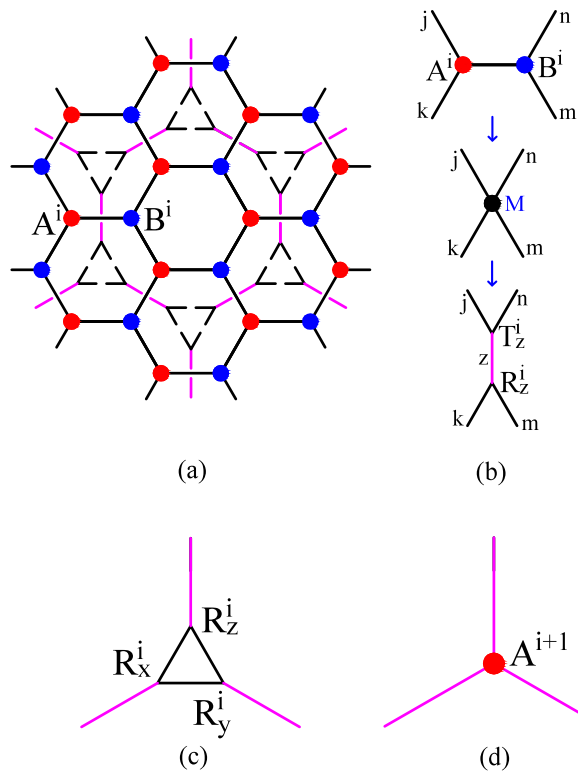}
\caption{ (color online) TRG coarse-graining scheme for honeycomb-lattice
tensor network\protect\cite{TRG2007}. Two sublattice tensor network is
assumed. (a) Rewire a honeycomb lattice to a triangle-honeycomb lattice. (b)
Schematic representation of Eq \protect\ref{eq:rewire2to1} and \protect\ref%
{eq:rewire1to2}. (c) Three i-th scale tensors are coarse grained to one
(i+1)-th scale tensor $A^{i+1}$ as in (d). }
\label{fig:honeycombTRG}
\end{figure}
%=========================

There are two steps at each coarse-graining
procedure. The first step is to rewire a honeycomb lattice to a
triangle-honeycomb lattice as shown in Fig. \ref{fig:honeycombTRG}(a). This
is done by contracting a pair of neighboring-site tensors on two sides of a
dashed line, which is explicitly described as
\begin{equation}
M_{jn,km}=\sum_{z}A_{zjk}^{i}B_{zmn}^{i}.  \label{eq:rewire2to1}
\end{equation}
and as shown pictorially in Fig. \ref{fig:honeycombTRG}(b). Then the SVD is
applied to decompose $M$ in the perpendicular direction and keep $\chi$
largest singular values
\begin{equation}
M_{jn,km}\approx\sum_{z=1}^{\chi}U_{jn,z}\Lambda_{z}V_{km,z},
\label{eq:rewire1to2}
\end{equation}
where truncation is performed and $\chi$ largest singular values are kept.
To estimate the quality of the truncation, we define the truncation error
as
\begin{equation}
\varepsilon=1-\frac{\sum_{z=1}^{\chi}\Lambda_{z}^{2}}{\sum_{z}\Lambda_{z}^{2}}.
\label{eq:truncErr}
\end{equation}

Then, the construction of $R_{z}^{i}$ and $T_{z}^{i}$ as shown pictorially in
the bottom of Fig. \ref{fig:honeycombTRG}(b) can be done as

\begin{equation}
R_{z}^{i}=U\Lambda^\frac{1}{2},\;
T_{z}^{i}=V\Lambda^\frac{1}{2}
\label{eq:rewireHoneycomb}
\end{equation}

The second step is to contract three tensors on every triangle in the
rewired lattice to form a honeycomb-lattice tensor network of which the size
is reduced by a factor of 3 (see Fig. \ref{fig:honeycombTRG}(c) to (d)). By
repeating this coarse-graining procedure, one can finally obtain a network
with six tensors on a hexagon which can be contracted exactly. The
contraction of the whole tensor network can then be achieved.

The computational cost for the TRG on the honeycomb lattices
scales as $O\left(\chi^{6}\right)$, because the most time consuming
computation is to apply SVD on $\chi^{2}\times \chi^{2}$ matrices.

%=========================
\begin{figure}[h]
\includegraphics[width=8.5cm]{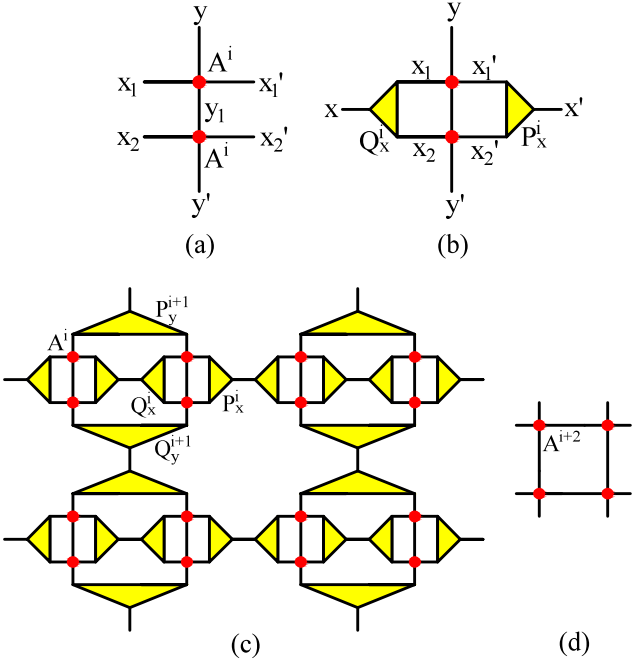}
\caption{(color online) HOTRG coarse-graining scheme for square-lattice
tensor network\protect\cite{HOTRG2012}. (a) two tensors aligned vertically
are contracted. (b) Construct projector $P_{x}^{i}$ and $Q_{x}^{i}$
illustrated by (yellow) triangles by HOSVD. (c) One vertical and one
horizontal coarse graining steps are performed, and four i-th scale tensors $%
A^{i}$ are coarse grained to one (i+2)-th scale tensor $A^{i+2}$ shown in
(d). }
\label{fig:HOTRG}
\end{figure}
%=========================

\subsection{Higher-order tensor renormalization group}

We next summarize the HOTRG procedure\cite{HOTRG2012}. During the TRG
procedure, the lattice shape changes alternatingly every after the scale
renormalization from the $i$-th to $i+1$-th scale, while HOTRG keeps lattice
shape during the procedure. For simplicity, we take a single-sublattice
square-lattice tensor network as an example to illustrate the HOTRG method.
At every coarse graining step, we contract 2 tensors to form one higher
scale tensor (Fig. \ref{fig:HOTRG}(b)), on which 2 projectors $P_{x}^{i}$,
and $Q_{x}^{i}$ are applied to make the truncation of the bond dimension.
Here we explain how to construct $P_{x}^{i}$ and $Q_{x}^{i}$. As shown in
Fig. \ref{fig:HOTRG}(a), two tensors aligned vertically are contracted
\begin{equation}
M_{\left(x_{1}x_{2}\right)\left(x_{1}^{\prime}x_{2}^{\prime}\right)yy^{%
\prime}}^{i}=\sum_{y_{1}}A_{x_{1}x_{1}^{\prime}yy_{1}}^{i}A_{x_{2}x_{2}^{%
\prime}y_{1}y^{\prime}}^{i}.
\end{equation}

In order to make the truncation, we apply the HOSVD on $M^{i}$ as\cite%
{HOTRG2012}
\begin{equation}
M_{\left(x_{1}x_{2}\right)\left(x_{1}^{\prime}x_{2}^{\prime}\right)yy^{%
\prime}}^{i}=\sum_{ijkl}S_{ijkl}U_{\left(x_{1}x_{2}\right)i}^{L}U_{%
\left(x_{1}^{\prime}x_{2}^{\prime}\right)j}^{R}U_{yk}^{U}U_{y^{\prime}l}^{D},
\end{equation}
where the sub-tensors in the core tensor $S$ are reordered so that the order
satisfies their Frobenius norm as %\begin{equation}
%\txrs{
%\left|S_{\ldots n\ldots}\right|\geq\left|S_{\ldots n^{\prime}\ldots}\right|,\qquad {\rm if}\quad n<n^{\prime},
%}
%\end{equation}
\begin{equation}
\quad n<n^{\prime}, \qquad \mathrm{if} \left|S_{\ldots
n\ldots}\right|\geq\left|S_{\ldots n^{\prime}\ldots}\right|,
\end{equation}

where $n$ and $n^{\prime}$ belong to the same index.

Then, we can make truncation by comparing
\begin{equation}
\varepsilon_{L}=\sum_{i>\chi}\left|S_{i,:,:,:}\right|^{2}
\end{equation}
and
\begin{equation}
\varepsilon_{R}=\sum_{j>\chi}\left|S_{:,j,:,:}\right|^{2}.
\end{equation}
if $\varepsilon_{L}<\varepsilon_{R}$, we keep the first $\chi$ columns in $U^{L}$
to form the truncated matrix $\tilde{U}^{L}$, thereby the projectors in the
horizontal direction can be constructed as
\begin{equation}
P_{x}^{i}=\tilde{U}^{L},\qquad Q_{x}^{i}=\left(\tilde{U}^{L}\right)^{%
\dagger}.  \label{eq:HOTRGprojector}
\end{equation}
If $\varepsilon_{L}\geq\varepsilon_{R}$, the way of constructing the projectors
follows Eq.~(\ref{eq:HOTRGprojector}) by replacing $\tilde{U}^{L}$ with $%
\tilde{U}^{R}$ which is obtained from the truncation of $U^{R}$.

After we obtain the projectors, we can contract 2 site tensors with
corresponding projectors to form a renormalized tensor network of which the
size is reduced by a factor of 2, and we can coarse grain the lattice
alternatively along the horizontal and vertical directions as shown in Fig. %
\ref{fig:honeycombTRG}(c)(d). The computational cost for this tensor
contraction scales as $O\left(\chi^{7}\right)$, which is the leading cost
for HOTRG on the square lattice.

Up to now, we have discussed the general procedure of TRG and HOTRG
algorithms to contract the tensor network. We note here that the partition
function and the expectation values of the physical observables can be
evaluated naturally in these coarse-graining procedure. To be concrete, let
us take HOTRG as an example. Suppose $\lambda_{i}$ is the renormalization
coefficient extracted (in order to keep the amplitude of any tensor element
within the digit allowed in computers) in the $i$-th iteration from a single
local tensor $T_{i}$. For example, $\lambda_i$ may be defined as the element
of $T_{i}$ with the largest absolute value before renormalizing the
amplitudes of the element of $T_i$. Then, the partition function can be
expressed as
\begin{align}
Z & =\mathrm{Tr}\{\prod_{i=1}^{N} T_{0}\}  \notag \\
& =\mathrm{Tr}\{\prod_{i=1}^{N/2}T_{1}\}\cdot\lambda_{1}^{N/2}  \notag \\
& =...  \notag \\
& =\mathrm{Tr}\{T_{\nu}\}\cdot\lambda_{1}^{N/2}\lambda_{2}^{N/2^{2}}...%
\lambda_{\nu}^{N/2^{\nu}}.
\end{align}
Here $N$ is the total number of tensors. If the system is reduced to
a single tensor $T_{\nu}$ after $\nu$ steps, then $N=2^{\nu}$. Actually this
is the effective formula we used in the practical calculation:
\begin{equation}
\frac{\ln Z}{N}=\sum_{i=1}^{\nu}\frac{1}{2^{i}}\ln\lambda_{i}+
\frac{\ln(\mathrm{Tr}%
\{T_{\nu}\})}{2^{\nu}}.
\end{equation}

For simplicity, a single sub-lattice structure is assumed. It seems that the
free energy is determined only by the coefficients $\lambda_{i}$s and the
final tensor $T_{\nu}$. In practical computations, in most cases, $\nu=40$
or so is sufficient to make the free energy reach the asymptotic value
expected in the thermodynamic limit. Similarly one can also calculate the
expectation value of any physical quantities. For a local operator $\hat{O}$%
, its expectation value can be expressed as
\begin{align}
\langle\hat{O}\rangle & =\frac{\mathrm{Tr}\{\hat{O}e^{-\beta H}\}}{\mathrm{Tr}%
\{e^{-\beta H}\}}  \notag \\
& =\frac{\mathrm{Tr}\{S_{0}\prod_{i=1}^{N-1} T_{0}\}}
{\mathrm{Tr}\{\prod_{i=1}^{N} T_{0}\}}  \notag \\
& =\frac{\mathrm{Tr}\{S_{1}\prod_{i=1}^{N/2-1}T_{1}\}}
{\mathrm{Tr}\{\prod_{i=1}^{N/2}T_{1}\}}  \notag \\
& =...  \notag \\
& =\frac{\mathrm{Tr}\{S_{n}\}}{\mathrm{Tr}\{T_{n}\}}
\label{eq:O}
\end{align}
where $S_{0}$ is a local tensor which contains the information of operator
and usually dubbed as an $impurity$ tensor. The denominator and numerator
are nothing but tensor networks, with the only difference that the numerator
has an impurity tensor. The central idea is to see how the impurity tensor
evolves in the coarse-graining process. After the contraction as in the
evaluation of the partition function, one can finally get the ratio. By
employing the same renormalization coefficients $\lambda_{i}s$ in each
iteration for the two networks (one for the denominator and the other for
the numerator), we obtain a practical formula to calculate the expectation
value of a single operator as in Eq.(\ref{eq:O}) after cancellation of $%
\lambda_i$s.

For two-body operators such as spin-spin correlation function, or other
complicated operators, the methods are quite similar, where the only difference
is the number of impurity tensors in the numerator.

\section{Second renormalization\label{sec:SRG}}

TRG and HOTRG are local optimization methods which optimize the
approximation of the local targeted system without any considerations of the
environment. In order to optimize the approximation of the whole tensor
network, it is necessary to consider the effect of environment tensor
network with respect to the tensors needed to be truncated. At the same
time, it is also necessary to calculate the environment tensor during the
imaginary time evolution. Therefore, we have developed SRG\cite%
{SRG2009,SRGfull2010} and HOSRG\cite{HOTRG2012} methods to consider the
renormalization effect of the environment and optimize the approximation in
the contraction of whole tensor networks. For the convenience of describing
finite size coarse-graining methods in the next section, we reiterate the
key procedures of SRG and HOSRG here.

\subsection{Second renormalization group}

%==================================================
\begin{figure}[tbp]
\includegraphics[width=8.5cm]{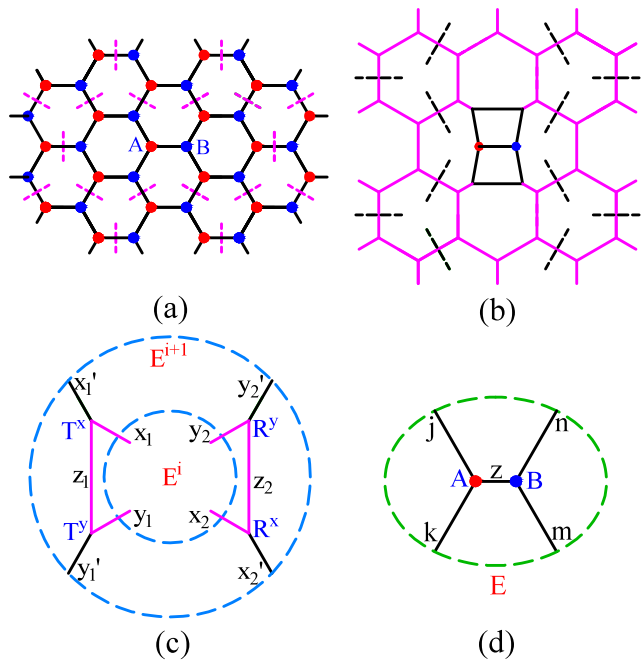}
\caption{(color online) SRG procedure for honeycomb lattice\protect\cite%
{SRG2009,SRGfull2010}. From (a) to (b) is one coarse-graining step for
environment sites. A pair of neighboring site tensors on two sides of a
dashed line are contracted and then decomposed along the dashed line.
(c) Relation between the neighboring scales environments $E^{i}$ and $E^{i+1}
$ (dashed circles) as represented in Eq.~(\protect\ref{eq:HoneyBackIter}).
(d) Environment tensor $E$ with system
site tensors $A$ and $B$. }
\label{fig:honeycombSRG}
\end{figure}
%==================================================

The general idea of SRG is to find an effective representation of the
environment by coarse graining the surrounding environment tensor network
with respect to the target tensors, and then use the information of the
environment to do global optimization in the approximation of the target
tensors. To be specific, let us explain how to implement the SRG on
the honeycomb lattice here.
The SRG method on the square lattice is described in Appendix \ref{apdx:squareSRG}.

Figure \ref{fig:honeycombSRG} shows the SRG procedure for the honeycomb
lattice\cite{SRG2009,SRGfull2010}.
First, we do forward iteration by the TRG, in order to contract the whole
surrounding tensor network. In this iteration, the local tensors at each
scale is stored.
A general guiding clue here is that one should make the
targeted system as compact as possible in the coarse-graining process.
As in Fig. \ref{fig:honeycombSRG},
we keep a pair of two neighboring sites
unchanged, and coarse grain the environment tensor network as shown in Figs.~%
\ref{fig:honeycombSRG}(a) and (b) by the TRG.
Since we
assume that the environment is always infinite at every coarse-graining step
of the system sites, we need to apply the coarse graining in the environment
for a sufficient number of steps until converged.

Second, we do backward iteration from the closing scale in order to find effective environment of the targeted system, represented also as a local tensor. In this iteration, the central problem is to find the recursion relation between the environment of two neighboring scales.
Figure \ref{fig:honeycombSRG}(c)
schematically shows the relation between the environments of the neighboring
scales,  which can be represented by the following iteration formula\cite%
{SRGfull2010}
\begin{equation}
E_{x_{1}y_{1}x_{2}y_{2}}^{i}=
E_{x_{1}^{\prime}y_{1}^{\prime}x_{2}^{\prime}y_{2}^{\prime}}^{i+1}
T_{z_{1}x_{1}x_{1}^{\prime}}^{x} T_{z_{1}y_{1}y_{1}^{\prime}}^{y}
R_{z_{2}x_{2}x_{2}^{\prime}}^{x} R_{z_{2}y_{2}y_{2}^{\prime}}^{y},
\label{eq:HoneyBackIter}
\end{equation}
where the repeated indices imply summation.

After a forward iteration, we are able to do backward iteration to contract
the network by starting from tracing out the outermost bonds and follow Fig.
\ref{fig:honeycombSRG}(c) to compute the environment tensor from higher to
lower scales. Eventually, we can obtain the
environment tensor $E=E^{1}$ which is defined as the outer part of the green
ellipse in Fig. \ref{fig:honeycombSRG}(d).

Then, we use the obtained environment tensor to do global optimization of
the local tensor (namely, the target tensor). Here the global optimization
means that the local truncation is chosen so that, compared to the untruncated case, the difference of the
partition function or the summation of the whole tensor network is minimized,
 as concretely detailed below.

After finding out the environment tensor $E$ (green dashed ellipse in Fig.
\ref{fig:honeycombSRG}(d)), we can contract the system sites $A$ and $B$
to
form a matrix $M$ as in Eq. \ref{eq:rewire2to1}.
The SVD is then applied to decompose $E$,
\begin{equation}
E_{km,jn}= \sum_{z} X_{km,z}\Omega_{z} Y_{jn,z},  \label{eq:SVDenv_honey}
\end{equation}
where $X$ and $Y$ are unitary matrices.

Then, we can take the environment contribution into account by constructing
the following tensor\cite{SRGfull2010}
\begin{equation}
\tilde{A}_{z_{1}z_{2}} =\sum_{mnjk} \Omega_{z_1}^{\frac{1}{2}} Y_{jn,z_{1}}
M_{jn,km} X_{km,z_{2}} \Omega_{z_{2}}^{\frac{1}{2}}.
\label{eq:sys_env_honey}
\end{equation}

We compute the SVD of $\tilde{A}$ and make truncation to keep $\chi$ largest
singular values $\Lambda$, the corresponding left singular vectors $U$ and
right singular vectors $V$ as
\begin{equation}
\tilde{A}_{z_{1}z_{2}} \approx \sum_{z=1}^{\chi} U_{z_{1}z}\Lambda_{z}
V_{z_{2}z}.
\label{eq:SVD_sysEnv}
\end{equation}

Finally, $M_{jn,km}$ in Eq.~(\ref{eq:rewire2to1}) can be truncated and decomposed
as a product of two tensors
\begin{equation}
M_{jn,km} \approx \sum_{z=1}^{\chi} R_{jnz} T_{kmz},  \label{eq:decompse_sys}
\end{equation}
where
\begin{equation}
R_{jnz} = \sum_{z_{1}} Y_{jn,z_{1}} \Omega_{z_{1}}^{-\frac{1}{2}} U_{z_{1}z}
\Lambda_{z}^{\frac{1}{2}}
\label{eq:R_honey}
\end{equation}
\begin{equation}
T_{kmz} = \sum_{z_{2}} X_{km,z_{2}} \Omega_{z_2}^{-\frac{1}{2}} V_{z_{2}z}
\Lambda_{z}^{\frac{1}{2}},
\label{eq:T_honey}
\end{equation}
are two rank-3 tensors similarly to Eq.~(\ref{eq:rewireHoneycomb}), which
are on two neighboring triangles connected by one bond.

Since the aim of the SRG is to optimize the truncation globally, the truncation error of the SRG can be defined as
\begin{equation}
\varepsilon=1-\frac{\sum_{mnjkz}E_{km,jn}R_{jnz}T_{kmz}}
{\sum_{mnjk}E_{km,jn}M_{jn,km}}.
\label{eq:SRGtruncErr}
\end{equation}
This definition is consistent with Eq.~(\ref{eq:truncErr}),
because if we take $E=M$,
Eqs. (\ref{eq:truncErr}) and (\ref{eq:SRGtruncErr}) are equivalent each other.

We can then contract three tensors on every triangle in the
rewired lattice to form one rank-3 tensor as a coarse-grained site on a
renormalized honeycomb-lattice tensor network as shown from Fig. \ref%
{fig:honeycombTRG}(c) to (d). Then, we can repeat this coarse-graining
procedure until the network can be contracted exactly.
Thus the contraction of the whole tensor network can be completed.

The computational cost of tensor contraction in backward iteration, which
is
shown in Fig. \ref{fig:honeycombSRG}(c) and Eq.~(\ref{eq:HoneyBackIter}),
scales as $O\left(\chi^{6}\right)$. This is the same complexity as the SVD,
so the computational cost for the SRG on the honeycomb lattice is $%
O\left(\chi^{6}\right)$.

%======================================================
\begin{figure}[tbp]
\includegraphics[width=8.5cm]{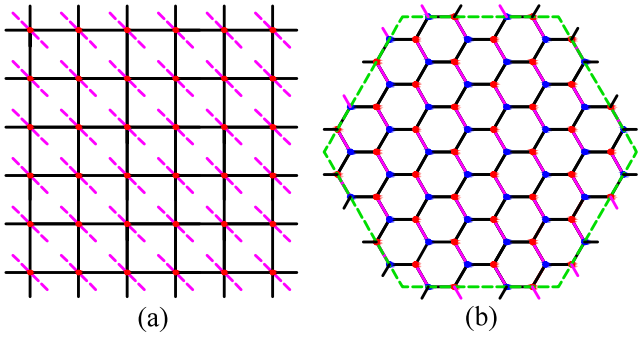}
\caption{(color online) Conversion from a square lattice to a honeycomb
lattice tensor network \protect\cite{SRGfull2010}. (a) Every square lattice
site tensor is decomposed by SVD along the direction indicated by dashed
(magenta) lines, and the $6\times6$ periodic boundary square lattice is
converted to (b) 72 sites honeycomb lattice with periodic boundary indicated
by green dashed hexagon. The parallel sides of the boundary are equivalent
sides. }
\label{fig:square2honeycomb}
\end{figure}
%======================================================

One can also apply the TRG to a square lattice directly with the computational cost scaling as $O\left(\chi^{10}\right)$, which is described in Appendix \ref{apdx:squareSRG}.
Fortunately, there exists a much cheaper method which needs to convert a
square lattice to a honeycomb lattice, and this can be done by SVD of every
fourth-order site tensor to two third-order tensors along the same diagonal
direction (Magenta dashed lines in Fig. \ref{fig:square2honeycomb}(a)) as
proposed in Ref. \onlinecite{SRGfull2010}.
After this conversion, we can
employ the honeycomb-lattice SRG method to contract this tensor network.
Obviously, this kind of conversion can be easily implemented in the
infinite-size lattice, because the boundary condition is not important in
the infinite system. For a finite square lattice with the PBC, this
conversion generates a honeycomb lattice with the hexagonal periodic
boundary, which means that the parallel sides of the boundary (green dashed
hexagon in Fig. \ref{fig:square2honeycomb}(b)) are equivalent sides. The
honeycomb lattice with hexagonal periodic boundary can be calculated by
TRG/SRG, and the size of lattice that can be handled is $(2\times3^{\frac{L}{%
2}})\times(2\times3^{\frac{L}{2}})$, where $L$ is the number of
coarse-graining steps.

\subsection{Higher-order second renormalization group}

\begin{figure}[tbp]
\includegraphics[width=8.5cm]{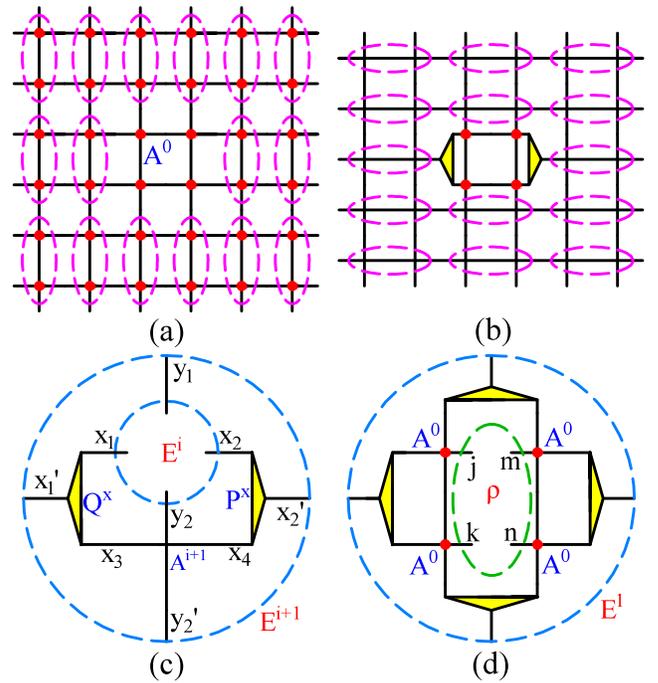}
\caption{(color online) HOSRG procedure for square lattice\protect\cite%
{HOTRG2012}. From (a) to (b) is one coarse-graining step for environment
sites. 4 site tensors in every dashed circle are coarse grained into 1
tensor. (c) Relation between the neighboring scales environments $E^{i}$
and
$E^{i+1}$ (dashed circles) as represented in Eq.~(\protect\ref%
{eq:backIter_HOSRG}). (d) Bond density matrix $\protect\rho$ can be obtained
by tracing out all bonds outside the green ellipse. }
\label{fig:squareHOSRG}
\end{figure}

The aim of the HOSRG is also to optimize the approximation of the whole
tensor network by considering the renormalization effect of the environment.
However, instead of the environment of site tensors, the environment of
bonds is computed. We keep $2$ bonds unchanged, and coarse grain the
environment tensor network as shown in Fig. \ref{fig:squareHOSRG}(a) and (b)
by the HOTRG.

%===========================================
%===========================================

Then, we can employ backward iteration to calculate the environment tensor.
Fig. \ref{fig:squareHOSRG}(c) schematically shows the relation between the
environments of the neighboring scales, which can be represented by the
following iteration formula\cite{HOTRG2012}
\begin{eqnarray}
& &
E_{x_{1}y_{1}x_{2}y_{2}}^{i}=E_{x_{1}^{\prime}y_{1}x_{2}^{\prime}y_{2}^{%
\prime}}^{i+1} P_{x_{2}^{\prime}x_{2}x_{4}}^{x}
Q_{x_{1}^{\prime}x_{1}x_{3}}^{x} A_{x_{3}x_{4}y_{2}y_{2}^{\prime}}^{i+1}.
\label{eq:backIter_HOSRG}
\end{eqnarray}

We start contracting the environment network by tracing out the outermost
bonds and follow Fig. \ref{fig:squareHOSRG}(c) to compute the environment
tensor from higher to lower scales. Eventually, we can obtain the bond
density matrix $\rho$ by tracing out all the bonds except the bonds to be
truncated, which is denoted as the bonds inside the green ellipse in Fig.
\ref{fig:squareHOSRG}(d).

$\rho$ is not always a symmetric matrix, so the diagonalization of $\rho$
may contain negative or complex eigenvalues. Instead, we construct the
projectors by a biorthonormal transformation\cite{Biorthonormal2011}. First,
We compute the SVD of $\rho$ and make truncation to keep $\chi$ largest
singular values $\Omega$, the corresponding left singular vectors $X$ and
right singular vectors $Y$ as
\begin{equation}
\rho_{jk,mn}\approx \sum_{x} X_{jk,x}\Omega_{x}Y_{mn,x}.  \label{eq:densMat}
\end{equation}
Then, we compute the SVD on $Y^{\dagger}X$
\begin{equation}
\sum_{jk} Y_{jk,x_{1}}X_{jk,x_{2}}= \sum_{x}U_{x_{1}x} \Lambda_{x}
V_{x_{2}x}.  \label{eq:biorth}
\end{equation}
Finally, we can construct the projectors
\begin{eqnarray}
P_{jk,x}^{x} & = & \sum_{x_{2}} X_{jk,x_{2}}V_{x_{2}x} \Lambda_{x}^{\frac{1}{%
2}} \\
Q_{mn,x}^{x} & = & \sum_{x_{1}} Y_{mn,x_{1}}U_{x_{1}x} \Lambda_{x}^{\frac{1}{%
2}}.  \label{eq:Projector}
\end{eqnarray}
The construction of projectors associated with vertical bonds is almost the
same, except that we need to compute the bond density matrix of vertical
bonds.

The computational cost for tensor contraction in backward iteration, which
is shown in Fig. \ref{fig:squareHOSRG}(c) and Eq.~(\ref{eq:backIter_HOSRG})
scales as $O\left(\chi^{7}\right)$, while the computational cost for the
construction of bond density matrix $\rho$ scales as $O\left(\chi^{8}\right)$%
, which is shown in Fig. \ref{fig:squareHOSRG}(d) Therefore, the leading
computational cost for HOSRG on the square lattice is $O\left(\chi^{8}\right)
$.

\section{Algorithm for finite periodic systems
\label{sec:finiteSRG}%
}

The SRG and HOSRG introduced in \ref{sec:SRG} are for the infinite lattices,
since at every coarse-graining step of system sites, the size of environment
tensor networks are always assumed to be infinite. In order to contract a
finite-lattice tensor network with the PBC by coarse-graining tensor
renormalization methods, we will introduce the implementation of SRG and
HOSRG on finite lattices. These finite lattice methods can be also applied
to the infinite lattice, since it further improves the accuracy by reducing
the accumulation error at every coarse-graining step.

Let us start by considering the finite-size SRG method on the honeycomb
lattice with the hexagonal PBC, and assume that the tensor network contains
two sublattices $A^{0}$ and $B^{0}$. Suppose that the honeycomb tensor
network contains $8\times3^{L}$ sites, where $L$ is the total number of the
coarse-graining steps. The following procedure is employed:

\begin{enumerate}
\item The TRG is employed to obtain and store site tensors $A^{i}$, $B^{i}$
( $i=0,\ldots,L$ ) and also $R_{\gamma}^{i}$, $T_{\gamma}^{i}$ ($%
i=0,\ldots,L-1$, $\gamma=x,y,z$) at every scale. $R_{\gamma}^{i}$, $%
T_{\gamma}^{i}$ are obtained from the transformation of $A^{i}$, $B^{i}$
along the $\gamma$ direction as shown in Fig. \ref{fig:honeycombTRG}(b).

\item For the $i$-th scale tensor network which contains $8\times3^{L-i}$
sites, we compute the environment of a pair of neighboring sites $A^{i}$, $%
B^{i}$ according to Fig. \ref{fig:honeycombSRG}. The number of
coarse-graining steps required to contract the environment tensor network is
$L-i$, and $i,\;i+1,\:\ldots,\:L$-th scale tensors $\left(R_{\gamma}^{i},T_{%
\gamma}^{i}\right),\left(R_{\gamma}^{i+1},T_{\gamma}^{i+1}\right),\ldots,%
\left(R_{\gamma}^{L-1},T_{\gamma}^{L-1}\right)$ and $\left(A^{L},B^{L}\right)
$ are needed to calculate the environment tensor $E$.

\item After the environment tensor $E$ is obtained, the environment
contribution can be taken into account by Eq.~(\ref{eq:SVDenv_honey}) and
Eq.~(\ref{eq:sys_env_honey}). Then, we can construct $R_{\gamma}^{i}$, $%
T_{\gamma}^{i}$ by Eq.~(\ref{eq:R_square}) and Eq.~(\ref{eq:T_square}),
respectively.

\item After all the $R_{\gamma}^{i}$, $T_{\gamma}^{i}$ updated in the same
way, we can contract $R_{z}^{i}$, $R_{x}^{i}$ and $R_{y}^{i}$ to form
updated $A^{i+1}$, and contract $T_{z}^{i}$, $T_{x}^{i}$ and $T_{y}^{i}$ to
form updated $B^{i+1}$ according to Fig. \ref{fig:honeycombTRG}(c)(d).

\item Since $A^{i+1}$, $B^{i+1}$ has been updated, tensors from $(i+2)$-th
scale to $L$-th scale should be changed accordingly before going to update
tensors of $(i+2)$-th scale. Here we use the TRG to obtain the $(i+2)$-th
scale to the $L$-th scale tensors which are required for coarse graining
from the $(i+1)$-th scale to $(i+2)$-th scale tensor network.

\item Repeat the steps 2 to 5 from $i = 0$ to $i = L - 1$.
\end{enumerate}

The difference from the SRG method for the infinite lattice is that the size
of the lattice is fixed and the number of tensors in the environment is
reduced in the process of coarse graining. In the original SRG method for
the infinite lattice, although the truncation error is reduced by
considering the environment contribution, the truncation error at every
coarse-graining step will accumulate. This accumulation appears unavoidable
for the infinite lattice algorithm. However, the finite-size algorithm has
its own advantage for the possibility of controlling and eliminating the
error accumulation. In order to reduce the accumulation error, we employ a
sweeping scheme, which is similar as the sweeping in finite size DMRG
algorithm. The sweeping scheme is to repeat the step 2 to step 6 until
converged. The computational cost for finite SRG scales as $%
O\left(N_{s}\cdot \chi^{6}\right)$, where $N_{s}$ is the number of sweeping
and $\chi^6$ is the cost for SRG calculation.

In the finite size SRG method described above, when the tensors at the $i$%
-th scale are updated by the SRG, all the tensors at higher scales are
updated simultaneously because of the step 5. As a matter of fact, if the
updating of $A^{i+1}$ and $B^{i+1}$ changes the tensors sufficiently slow,
we can assume that the environment of $A^{i+1}$ and $B^{i+1}$ does not
change. Then, we can skip step 5 to reduce the computation cost.

The finite-size HOSRG method is very similar to the finite-size SRG method,
so we only explain the difference of them. Let us consider a square lattice
with the PBC. Suppose that the tensor network contains $4\times2^{L}$ sites,
where $L$ is the total number of coarse-graining steps. In the step 1 and
step 5, the TRG is replaced by the HOTRG. Since the HOTRG/HOSRG scheme does
not contain the lattice rewiring, $R_{\gamma}^{i}$, $T_{\gamma}^{i}$ do not
exist. Instead, we need to store all the projectors $P_{\gamma}^{i}$, $%
Q_{\gamma}^{i}$ ($i=0,\ldots,L-1$, $\gamma=x,y$).

\section{Results\label{sec:benchmark}}

\subsection{Ising model on square lattice}

To benchmark our finite-size algorithms, we first calculate the Ising model
on the square lattice with the PBC. The partition function can be exactly
evaluated on any lattice size\cite{finiteIsing1949}. We first convert the
square-lattice tensor network to that of the honeycomb lattice as shown in
Fig. \ref{fig:square2honeycomb}. Then, we apply a finite SRG on the
honeycomb lattice to contract the tensor network, so the computational cost
scales as $O\left(\chi^{6}\right)$. In this calculation, we find that the
result is almost unchanged after the sweeping procedure. The reason is that
the truncation error of the finite SRG on this model is sufficiently small
and the number of the required coarse graining is also small, so that the
accumulated error does not practically cause a damage in the final result
even without the sweeping thanks to the simplicity of the model.

%=============================================
\begin{figure}[tbp]
\includegraphics[width=8.5cm]{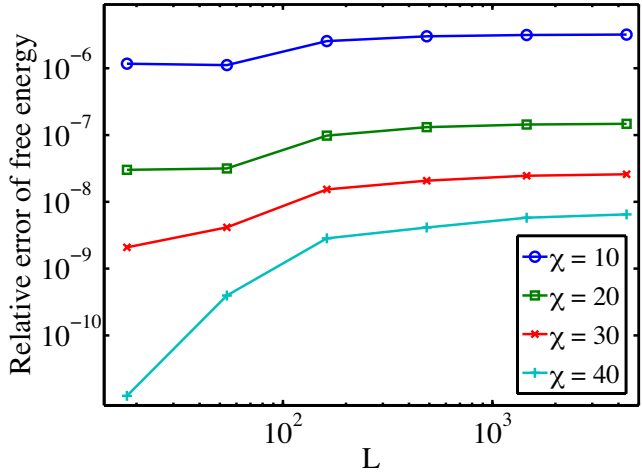}
\caption{(color online) Relative errors of free energy as a function of $L$
for $L \times L$ square-lattice Ising model computed by the SRG method for
finite lattices with $\protect\chi=10,\,20,\,30,\, 40$. The calculation is
at the critical temperature of the infinite lattice, namely at the
temperature $T=T_{c}=2/\ln\left(1+\protect\sqrt{2}\right)$. The exact free
energy for $18\times18$, $54\times54$, $162\times162$, $486\times486$, $%
1458\times1458$, and $4374\times4374$ are $-2.114134648928$, $-2.110149135739
$, $-2.109706474810$, $-2.109657292382$, $-2.109651827694$, and $%
-2.109651220507$, respectively.}
\label{fig:lnZerror_chi}
\end{figure}
%=============================================

\begin{figure}[tbp]
\includegraphics[width=8.5cm]{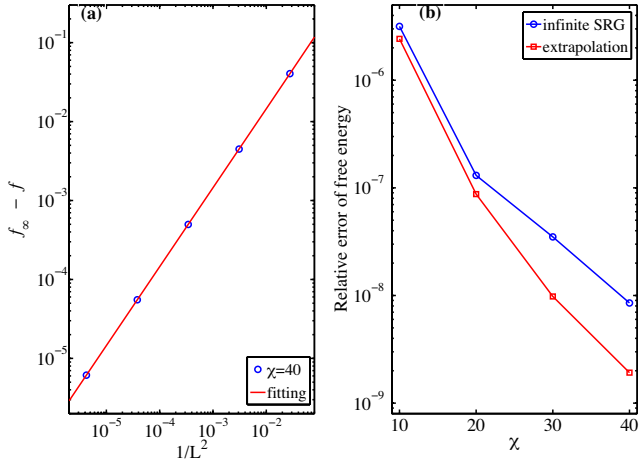}
\caption{(color online) (a) $(f_{\infty}-f\left( L \right))$ as a function
of $1/L^2$ (blue circles) at $T_{c}=2/\ln\left(1+\protect\sqrt{2}\right)$,
where $f\left( L \right)$ is the free energy. The red line shows the fit to
the function $f_{\infty}-f\left( L \right)=a/L^2+b/L^4$. (b) Relative errors
of free energy as a function of $\protect\chi$ computed by infinite SRG
(blue line with circles) and by finite SRG with extrapolation to infinite
size (red line with squares) by following the procedure in (a). }
\label{fig:finite_infinite}
\end{figure}

Figure \ref{fig:lnZerror_chi} shows the relative error of the free energy as
a function of $L$ for the Ising model on the $L \times L$ square lattice at
the critical temperature of the infinite lattice, $T_{c}=2/\ln\left(1+\sqrt{2%
}\right)$. We can see that the relative error basically increases with the
increase in the lattice size. This general trend applies to the
infinite-size calculation, where the error is the largest if one fixes $\chi$%
. Therefore, by applying the finite-size scaling to obtain infinite-size
free energy as shown in Fig. \ref{fig:finite_infinite}, the extrapolated
infinite-size result becomes more accurate than that of infinite-size SRG
calculation. In the extrapolation, we utilize the known finite-size scaling
of free energy with respect to $1/L^2$ at the critical temperature to the
second order\cite{IsingScaling2002}
\begin{equation}
f_{\infty}-f\left( L \right)= a/L^{2}+b/L^{4}+O\left( L^{6} \right),
\label{eq:finiteScaling}
\end{equation}
where $f_{\infty}$ is the extrapolated infinite-size free energy.

\begin{figure}[tbp]
\includegraphics[width=8.5cm]{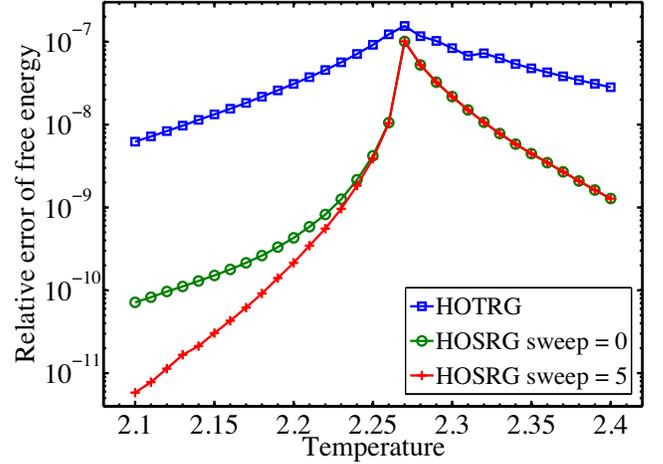}
\caption{(color online) Relative errors of free energy as a function of
temperature for the square-lattice Ising model by the HOTRG (square), the
HOSRG at the size $2^{25}\times2^{25}$ without (open circle) and with
(cross) 5 times sweeping . All the results are obtained with $\protect\chi=20
$.}
\label{fig:lnZerror_HOSRG}
\end{figure}

We also apply the finite HOSRG method to study the square-lattice Ising
model. When the lattice size is similar {to that in the} SRG calculation
above, typically, up to 10$^4$ in the linear dimension $L$, we also find
that the sweeping does not improve the accuracy significantly in this simple
model. However, when we increase the number of coarse-graining steps, namely
further increase the system size beyond $L=10^4$, the accumulated error
becomes more and more significant so that the sweeping provides us with a
substantially better accuracy. Figure \ref{fig:lnZerror_HOSRG} shows the
relative error of the free energy as a function of temperature for the $%
2^{25}\times2^{25}$ square lattice, which can be practically regarded as the
infinite lattice. We find that the sweeping scheme improves the accuracy
below the critical temperature. Therefore, even when one computes properties
in the thermodynamic-limit, our finite coarse-graining methods offers a
higher accuracy than the corresponding infinite-lattice methods.

\subsection{Kitaev model on honeycomb lattice}

Next, we calculate the Kitaev model on the honeycomb lattice\cite{Kitaev2006}
with an equal amplitude of bond couplings defined by
\begin{equation}
H=\frac{1}{2}\sum_{\left\langle i,j\right\rangle
_{\gamma}}\sigma_{i}^{\gamma}\sigma_{j}^{\gamma},\quad\gamma=z,x,y,
\end{equation}
where the Ising anisotropy depends on the three different directions of
bonds of the honeycomb lattice. Here, $\sigma_i^{\gamma}$ represents the
Ising spin at $i$ site with the Ising anisotropy axis in the $\gamma$
direction. This model is exactly solvable at any lattice size with any kind
of the PBC\cite{Kitaev2006}. The exact ground state is a highly frustrated
spin liquid and it is gapless in the thermodynamic limit.
The infinite PEPS calculation\cite{KHiPEPS2014} shows good agreement with the exact ground state energy in the infinite lattice, although the lattice rotational symmetry is artificially broken due to the mapping from honeycomb lattice to brick-wall lattice.

In our calculation, the wave function is represented as a honeycomb PEPS with the hexagonal
shaped PBC, and the imaginary time evolution is employed to determine the
ground state. As described in Sec. \ref{sec:ConstructTN}, the computation of
the environment is needed to optimize tensors during imaginary time
evolution. We apply the SRG method formulated for finite size to contract
the environment tensor network. We start the evolution from a relatively
large $\tau$ and gradually reduce the value of $\tau$ until converged. In
this calculation, we reduce $\tau$ from $0.1$ to $0.01$, since the ground
state energy does not show significant change any more when $\tau$ is
smaller than $0.01$.

After obtaining the ground-state wave function, we apply the SRG method for
finite size to compute the energy, the algorithm of which has also been
described in Sec. \ref{sec:ConstructTN}. In this calculation, the six-fold
rotational symmetry is always preserved in the coarse-graining procedure.

While the PEPS wave function itself satisfies the variational principle, the
computation of the expectation value by an approximate tensor-network
contraction does not necessarily satisfy the variational principle because
of the approximate contraction. In fact, the norm $\left\langle
\Psi|\Psi\right\rangle$ is calculated by the contraction of a tensor network
which is obtained by tracing out the physical indices in the PEPS wave
function, and the approximate contraction of this tensor network is not
guaranteed to be positive. As a result, the energy calculated approximately
from the PEPS wave function can be lower than the exact ground-state energy.
Therefore, it is necessary to assure the convergence in the computation of
the expectation value under a given PEPS wave function by improving the
truncation.

\begin{figure}[tbp]
\includegraphics[width=8.5cm]{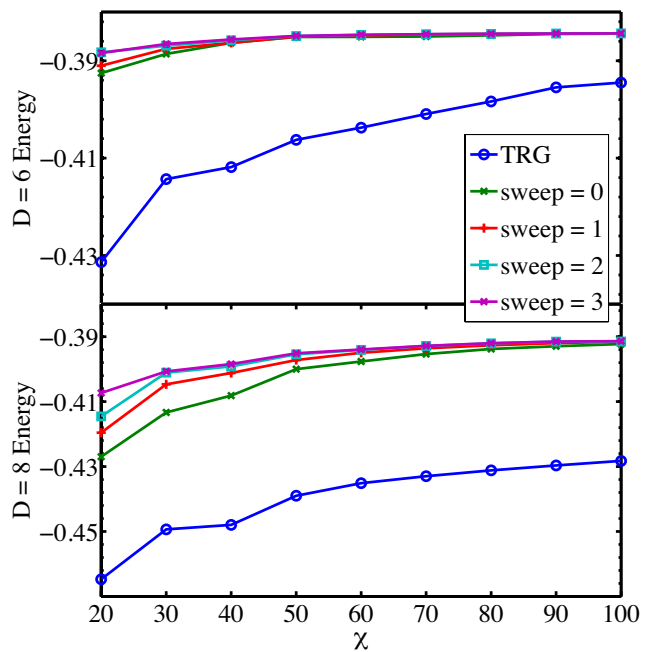}
\caption{(color online) Ground-state energy as a function of $\protect\chi$
for Kitaev model on honeycomb lattice with 216 sites. The upper panel shows $%
D=6$ results and the lower panel shows the $D=8$ results. The TRG, finite
SRG without sweeping and finite SRG with 1, 2, and 3 times sweeping results
are shown in both figures. The exact energy is -0.393752537.}
\label{fig:E_chi_L18L6M0}
\end{figure}

Figure \ref{fig:E_chi_L18L6M0} shows the convergence of the ground-state
energy with increasing $\chi$ for the Kitaev model on the 216-site lattice
with the wave-function tensor dimension $D=6$ (upper panel) and $D=8$ (lower
panel).
Our calculation shows that the truncation error in the TRG is quite large,
and the SRG can significantly reduce the truncation error.
In the case of $D=6$, and $\chi=50$,
the truncation error in the TRG is about $50\%$,
while the truncation error in the SRG can be reduced to the order of $10^{-3}$.
As a result, we can see that nevertheless the energy of the TRG is far away from the converged value,
the SRG calculation for the finite-size lattice makes a significant improvement.
While the energy computed from the
TRG calculation never get converged for $D=6$ and $D=8$ with $\chi$ up to
100, the SRG for the finite-size calculation has converged when $\chi>50$
for $D=6$ and $\chi>70$ for $D=8$.
We see that the sweeping procedure gives
more significant improvement on $D=8$ than $D=6$. That is because the
truncation error increases with increasing $D$ at a given $\chi$, indicating
that the sweeping becomes more important in reducing the truncation error
and in achieving a quicker convergence with relatively small $\chi$ when $D$
becomes large.

\begin{figure}[tbp]
\includegraphics[width=8.5cm]{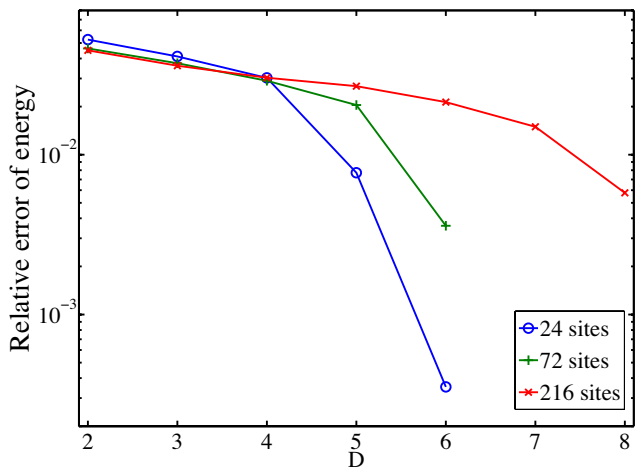}
\caption{(color online) Relative error of ground state energy as a function
of $D$ for Kitaev model on 24, 72, and 216 sites honeycomb lattice with
hexagonal shaped periodic boundary condition. Finite SRG method is applied
with three-times sweeping and $\protect\chi$ is taken up to 100 to verify
the convergence. }
\label{fig:error_D}
\end{figure}

Figure \ref{fig:error_D} shows the convergence of the ground-state energy
with increase of the bond dimension $D$ on several choices of lattice size.
The results are obtained by the finite SRG method with three-times sweeping.
We observe that the energy error continues to decrease rapidly and
unlimitedly with increasing $D$, indicating that the accuracy can be
improved without bounds with the increase of $D$.

\section{Summary \label{sec:summary}}

In this paper, We have developed coarse-graining tensor renormalization
group methods on finite periodic lattices and examined the efficiency of
these methods. The TRG and HOTRG are local optimization methods which
minimize the truncation error of the local tensors. In order to optimize the
approximation of the whole tensor network, it is necessary to consider the
effect of the environment tensor network with respect to the tensors needed
to be truncated. Therefore, SRG and HOSRG methods are introduced to contract
environment tensor networks and optimize the truncation of corresponding
site tensors.

We have also described the algorithms and prescriptions and the way how to
implement SRG and HOSRG methods to compute tensor network for finite
periodic lattices. In this algorithm, differently from that for the infinite
lattice, the environment tensor network is calculated with the size of
lattice at specific coarse-graining scales. When the truncation error is
large, we apply the sweeping scheme to further improve the accuracy. This
sweeping enables an accurate contraction of the tensor network beyond the
infinite-size algorithm but with the computational complexity similar to the
corresponding infinite lattice methods. After obtaining accurate finite-size
results, a reliable finite-size scaling allows to reach the thermodynamic
properties, that is more accurate than that obtained from the corresponding
infinite lattice method.

During the calculation, we use tensor networks with translational invariance
to perform the calculation. However, it is straightforward to implement the
calculation on systems without translational invariance, such as spin glass
systems. The difference there is that one
needs to calculate respective environment depending on every system site.

\section{ACKNOWLEDGMENTS}

We would like to thank Sotoshi Morita for providing us a python script to
calculate exact energy of finite lattice Kitaev model. The authors thank the
Supercomputer Center, the Institute for Solid State Physics, the University
of Tokyo for the facilities. This work was financially supported by the MEXT
HPCI Strategic Programs for Innovative Research (SPIRE) and the
Computational Materials Science Initiative (CMSI) under the project number
hp130007 and hp140215, and the National Natural Science Foundation of
China(Grants No.10934008, No.10874215, and No.11174365).

\appendix
\section{The square lattice TRG method\label{apdx:squareTRG}}

In this appendix we reiterate the procedure of TRG on the square lattice, which was proposed in Ref. \onlinecite{TRG2007}.

Here the single-sublattice
tensor network is taken as an example. Every
coarse-graining procedure contains two steps. The first step is to apply
SVD
to decompose every rank-4 site tensor
\begin{equation}
A_{jk,mn}^{i}\approx\sum_{x=1}^{\chi}U_{jk,x}\Lambda_{x}V_{mn,x},
\label{eq:SVDsquare}
\end{equation}
where truncation is performed and $\chi$ largest singular values are kept.

\begin{figure}[tbp]
\includegraphics[width=8.5cm]{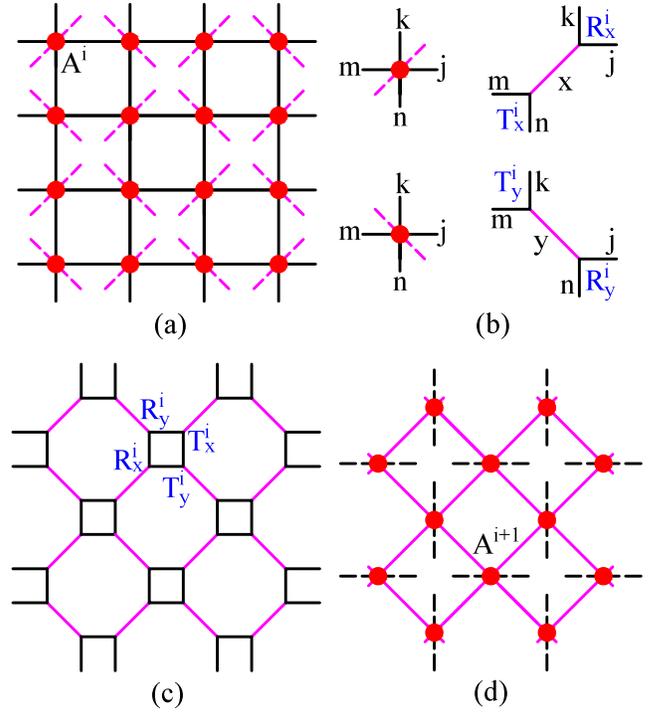}
\caption{(color online) TRG coarse-graining scheme for square lattice tensor
network\protect\cite{TRG2007}. two-sublattice tensor network is assumed.
(a)
Dashed lines show SVD directions. (b) SVD to decompose every rank-4 site
tensor to two rank-3 tensors and keep $\protect\chi$ largest singular
values. (c) Tensor network structure after SVD. (d) is formed from
contraction of 4 tensors on every black square in (c), and the green dashed
lines are the SVD directions for the next step coarse graining. }
\label{fig:squareTRG}
\end{figure}

The direction of SVD is depicted as dashed lines in Fig. \ref{fig:squareTRG}%
(a). Then, 2 new tensors
\begin{equation}
R_{x}^{i}=U\Lambda^\frac{1}{2},\; T_{x}^{i}=V\Lambda^\frac{1}{2}
\label{eq:rewireSquare}
\end{equation}
are formed as shown in the top of Fig. \ref{fig:squareTRG}(b). $R_{y}^{i}$
and $T_{y}^{i}$ shown in the bottom of Fig. \ref{fig:squareTRG}(b) are
constructed similarly as $R_{x}^{i}$ and $T_{x}^{i}$ by Eq.~(\ref%
{eq:SVDsquare}) and Eq.~(\ref{eq:rewireSquare}) except that the SVD is
applied in the perpendicular direction.

Then, the network changes the shape to Fig. \ref{fig:squareTRG}(c), and we
can contract four rank-3 tensors on every black square to form one rank-4
tensor as a coarse-grained site and then form a square-lattice tensor
network, of which the size is reduced by a factor of two with the 45-degrees
rotated lattice configuration (Fig. \ref{fig:squareTRG}(d)). By repeating
this coarse-graining procedure, one can finally obtain a 2x2 square-lattice
tensor network, which can be contracted exactly. Then the contraction of
the
whole tensor network is completed.

%=========================
%=========================

\section{The square lattice SRG method\label{apdx:squareSRG}}

In this appendix, we introduce how to implement the SRG on
the square lattice.

%======================================================
\begin{figure}[tbp]
\includegraphics[width=8.5cm]{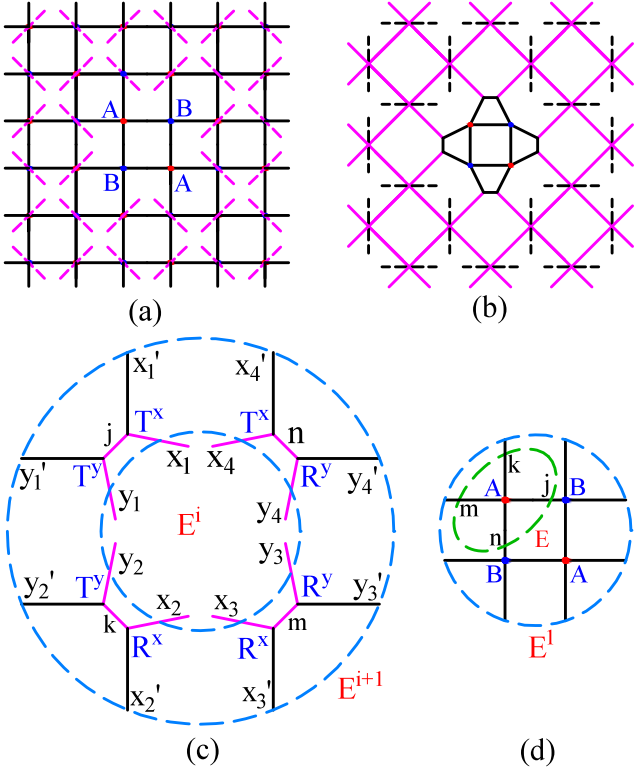}
\caption{(color online) SRG procedure for square lattice. From (a) to (b)
is
one coarse-graining step for environment sites. The site tensors in the
environment are decomposed along the dashed lines. (c) Relation between the
the neighboring scales environments $E^{i}$ and $E^{i+1}$ (dashed circles)
as represented in Eq.~(\protect\ref{eq:backIter}). (d) Environment tensor
$E$
of one system site tensor $A$ is obtained by tracing out the outer part of
the green ellipse. }
\label{fig:squareSRG}
\end{figure}
%======================================================

First, we do forward iteration by the TRG to contract the whole
surrounding tensor network. In this iteration, the local tensors at each
scale is stored. As in Fig. \ref{fig:squareSRG}, we keep four
sites on a plaquette unchanged, and coarse grain the environment tensor
network as shown in Fig. \ref{fig:squareSRG}(a)(b)
by the square lattice TRG.
Since we assume that the environment is always infinite at every coarse-graining step
of the system sites, we need to apply the coarse graining in the environment
for a sufficient number of steps until converged.

Second, we do backward iteration to find the environment tensor E.
Figure
\ref{fig:squareSRG}(c) schematically shows the relation between the
environments of the neighboring scales, which can be represented by the
following iteration formula
\begin{eqnarray}
& & E_{x_{1}y_{1}x_{2}y_{2}x_{3}y_{3}x_{4}y_{4}}^{i}  \notag \\
& = & E_{x_{1}^{\prime}y_{1}^{\prime}x_{2}^{\prime}
y_{2}^{\prime}x_{3}^{\prime} y_{3}^{\prime}x_{4}^{\prime}
y_{4}^{\prime}}^{i+1}T_{jx_{1}x_{1}^{\prime}}^{x}
T_{nx_{4}x_{4}^{\prime}}^{x}  \notag \\
& & T_{jy_{1}y_{1}^{\prime}}^{y}T_{ky_{2}y_{2}^{\prime}}^{y}
R_{kx_{2}x_{2}^{\prime}}^{x}R_{mx_{3}x_{3}^{\prime}}^{x}
R_{my_{3}y_{3}^{\prime}}^{y}R_{ny_{4}y_{4}^{\prime}}^{y},
\label{eq:backIter}
\end{eqnarray}
where the repeated indices imply summation.

After a forward iteration, we are able to do backward iteration to contract
the network by starting from tracing out the outermost bonds and follow Fig.
\ref{fig:squareSRG}(c) to compute the environment tensor from higher to
lower scales. Eventually, we can calculate the environment tensor $E$, which
is defined as the outer part of the green ellipse in Fig. \ref{fig:squareSRG}%
(d), from $E^{1}$. Note that $E$ contains only 4 indices as $E_{jk,mn}$,
while every $E^{i}$ contains 8 indices.

After finding out the environment tensor $E$, SVD is applied to decompose
$E$
\begin{equation}
E_{jk,mn}=\sum_{z} X_{jk,z}\Omega_{z} Y_{mn,z}.  \label{eq:SVDenv_square}
\end{equation}

Then, we can take the environment contribution into account by constructing
the following tensor
\begin{equation}
\tilde{A}_{z_{1}z_{2}} =\sum_{mnjk}\Omega_{z_1}^{\frac{1}{2}} Y_{mn,z_{1}}
A_{mn,jk} X_{jk,z_{2}} \Omega_{z_{2}}^{\frac{1}{2}}.  \label{eq:sys_env}
\end{equation}

We compute the SVD of $\tilde{A}$ and make truncation to keep $\chi$ largest
singular values $\Lambda$, the corresponding left singular vectors $U$ and
right singular vectors $V$ as
\begin{equation}
\tilde{A}_{z_{1}z_{2}} \approx \sum_{z=1}^{\chi} U_{z_{1}z}\Lambda_{z}
V_{z_{2}z}.  \label{eq:SVD_sysEnv_square}
\end{equation}

Finally, the system site tensor $A_{mn,jk}$ can be truncated and decomposed
as a product of two tensors
\begin{equation}
A_{mn,jk} \approx \sum_{z=1}^{\chi} R_{mnz} T_{jkz},  \label{eq:decompse_sys_square}
\end{equation}
where
\begin{equation}
R_{mnz} = \sum_{z_{1}} Y_{mn,z_{1}} \Omega_{z_{1}}^{-\frac{1}{2}} U_{z_{1}z}
\Lambda_{z}^{\frac{1}{2}}  \label{eq:R_square}
\end{equation}
\begin{equation}
T_{jkz} = \sum_{z_{2}} X_{jk,z_{2}} \Omega_{z_2}^{-\frac{1}{2}} V_{z_{2}z}
\Lambda_{z}^{\frac{1}{2}},  \label{eq:T_square}
\end{equation}
are two rank-3 tensors as similar as in Eq.~(\ref{eq:rewireSquare}), which
are on two neighboring black squares connected by a (magenta) bold bond
(note that 45$^\circ$ tilted) in Fig. \ref{fig:squareTRG}(c). Then, the
network changes the shape to Fig. \ref{fig:squareTRG}(c), and we can
contract four rank-3 tensors on every black square to form one rank-4 tensor
as a coarse-grained site and then repeat this coarse-graining procedure
until the network can be contracted exactly. Thus the contraction of the
whole tensor network can be completed.

The leading computational cost for the above method with the direct coarse
graining on the square lattice is $O\left(\chi^{10}\right)$, because the
computational cost of backward iteration to contract the environment tensor
$%
E$, which is shown in Fig. \ref{fig:squareSRG}(c)(d) and Eq.~(\ref%
{eq:backIter}), scales as $O\left(\chi^{10}\right)$.


\begin{thebibliography}{99}
\bibitem{Niggemann1997} H. Niggemann, A. Klumper, and J. Zittartz, Z. Phys.
B \textbf{104}, 103 (1997).

\bibitem{PTP2001Nishino} T. Nishino, Y. Hieida, K. Okunishi, N. Maeshima, Y.
Akutsu, and A. Gendiar,  Prog. Theor. Phys. \textbf{105}, 409 (2001).

\bibitem{PEPS2004} F. Verstraete and J. I. Cirac, arXiv:cond-mat/0407066.

\bibitem{MERA} G. Vidal, Phys. Rev. Lett. \textbf{99}, 220405 (2007).

\bibitem{TRG2007} M. Levin and C. P. Nave, Phys. Rev. Lett. \textbf{99},
120601 (2007).

\bibitem{PRL2007Vidal} G. Vidal, Phys. Rev. Lett. \textbf{98}, 070201 (2007)

\bibitem{SimpleUpdate2008} H. C. Jiang, Z. Y. Weng, and T. Xiang, Phys. Rev.
Lett. \textbf{101}, 090603 (2008).

\bibitem{TEFR} Z. C. Gu, and X. G. Wen, Phys. Rev. B \textbf{80}, 155131
(2009).

\bibitem{SRG2009} Z. Y. Xie, H. C. Jiang, Q. N. Chen, Z. Y. Weng, and T.
Xiang, Phys. Rev. Lett. \textbf{103}, 160601 (2009).

\bibitem{SingleLayer2011} Iztok Pirn, Ling Wang, and Frank
Verstraete, Phys. Rev. A \textbf{83}, 052321 (2011).

\bibitem{HOTRG2012} Z. Y. Xie, J. Chen, M. P. Qin, J. W. Zhu, L. P. Yang,
and T. Xiang, Phys. Rev. B \textbf{86}, 045139 (2012).

\bibitem{PESS2014} Z. Y. Xie, J. Chen, J. F. Yu, X. Kong, B. Normand,
and T. Xiang, Phys. Rev. X \textbf{4}, 011025 (2014).

\bibitem{TNR} G. Evenbly, and G. Vidal, arXiv:1412.0732.

\bibitem{Book1982Baxter} R. J. Baxter, Exactly Solved Models in Statistical
Mechanics (Academic Press, London, 1982).

\bibitem{PRE2008Berker} M. Hinczewski, and A. N. Berker,
Phys. Rev. E \textbf{77}, 011104 (2008).

\bibitem{SRGfull2010} H. H. Zhao, Z. Y. Xie, Q. N. Chen, Z. C. Wei, J. W.
Cai, and T. Xiang, Phys. Rev. B \textbf{81}, 174411 (2010).

\bibitem{PRE2014Yujifeng} J. F. Yu, Z. Y. Xie, Y. Meurice, Y. Z. Liu, A.
Denbleyker, H. Y. Zou, M. P. Qin, J. Chen, and T. Xiang,
Phys. Rev. E \textbf{89}, 013308 (2014).

\bibitem{PRD2014Alan} A. Denbleyker, Y. Z. Liu, Y. Meurice, M. P. Qin, T.
Xiang, Z. Y. Xie, J. F. Yu, and Haiyuan Zou,
Phys. Rev. D \textbf{89}, 016008 (2014).

\bibitem{AKLT} I. Affleck, T. Kennedy, E. H. Lieb, and H. Tasaki,
Phys. Rev. Lett. \textbf{59}, 799 (1987);
Commun. Math. Phys. \textbf{115}, 477 (1988).

\bibitem{Sierra1998} G. Sierra, and M. Martin-Delgado, Proceedings of the
workshop on the Exact Renormalization Group, Faro(Portugal),  10-12
September 1998 (World Scientific, Singapore, in press).

\bibitem{RMP2010Eisert} J. Eisert, M. Cramer, and M. B. Plenio,
Rev. Mod. Phys. \textbf{82}, 277 (2010).

\bibitem{CPL2014Wang} S. Wang, Z. Y. Xie, J. Chen, B. Normand, and T. Xiang, Chin. Phys. Lett. \textbf{31}, 070503 (2014).

%\bibitem{CMP2012White} E. M. Stoudenmire, and S. R. White, Annu. Rev. Cond. %Matt. Phys. \textbf{3}, 111 (2012).

\bibitem{PRL2010Cirac} F. Verstraete, J. I. Cirac, Phys. Rev. Lett.
\textbf{104}, 190405 (2010).

\bibitem{PRB2014Zueco} F. Quijandria, J. J. Garcia-Ripoll, and D. Zueco,
Phys. Rev. B \textbf{90}, 235142 (2014).

\bibitem{PRA2010Cirac} C. V. Kraus, N. Schuch, F. Verstraete, and J. I.
Cirac, Phys. Rev. A \textbf{81}, 052338 (2010).

\bibitem{arXiv2010Wen} Z. C. Gu, F. Verstraete,
and X. G. Wen, arXiv:1004.2563.

\bibitem{PRB2015Kao} O. Sikora, H.-Wen Chang, C.-Pin Chou, F. Pollmann, and
Y.-Jer Kao, Phys. Rev. B \textbf{91}, 165113 (2015).

\bibitem{arXiv2015Eisert} C. Krumnow, O. Legeza, and J. Eisert,
arXiv:1504.00042.

\bibitem{PRB2014Zhou} C. Wang, S. M. Qin, and H. J. Zhou, Phys. Rev. B
\textbf{90}, 174201 (2014).

\bibitem{JPSJ1995Nishino} T. Nishino, J. Phys. Soc. Jpn. \textbf{64}, 3598
(1995).

\bibitem{TMRG1996} R. Bursill, Tao Xiang, and G. A. Gehring, J. Phys.: Cond.
Matter \textbf{8}, L583 (1996).

\bibitem{PRB1997Xiang} X. Q. Wang, and T. Xiang, Phys. Rev. B \textbf{56},
5061 (1997).

\bibitem{MP1968Baxter} R. J. Baxter, J. Math. Phys. \textbf{9}, 650 (1968).

\bibitem{NishinoCTMRG1996} T. Nishino and K. Okunishi, J. Phys. Soc. Jpn.
\textbf{65}, 891 (1996).

\bibitem{OrusCTMRG2009} Roman Orus and Guifre Vidal,
Phys. Rev. B \textbf{80}, 094403 (2009).

\bibitem{PRB2010Corboz} P. Corboz, J. Jordan, and G. Vidal, Phys. Rev. B
\textbf{82}, 245119 (2010).

\bibitem{QIC2007Verstraete} D. P.-Garcia, F. Verstraete, M. M. Wolf,
and J. I. Cirac, Quan. Inf. Comp. \textbf{7}, 401 (2007).

\bibitem{canonical2008} R. Orus and G. Vidal, Phys. Rev. B
\textbf{78}, 155117 (2008).

\bibitem{TERG2008} Z. C. Gu, M. Levin, and X. G. Wen,
Phys. Rev. B \textbf{78}, 205116 (2008).

\bibitem{PRB2014Nishino} H. Ueda, K. Okunishi, and T. Nishino,
Phys. Rev. B \textbf{89}, 075116 (2014).

\bibitem{IsingScaling2002} N. Sh. Izmailian, and Chin-Kun Hu,
Phys. Rev. E \textbf{65}, 036103 (2002).

\bibitem{PEPS2007} V. Murg, F. Verstraete, and J. I. Cirac, Phys. Rev. A
\textbf{75}, 033605 (2007).

\bibitem{PEPS2008} F. Verstraete, V. Murg, and J. I. Cirac, Adv. Phys. 57,
143 (2008).

\bibitem{iPEPS2008} J. Jordan, R. Orus, G. Vidal, F. Verstraete, and
J. I. Cirac, Phys. Rev. Lett. \textbf{101}, 250602 (2008).

\bibitem{finitePEPS2014} M. Lubasch, J. I. Cirac, and M-C Banuls,
Phys. Rev. B \textbf{90}, 064425 (2014).

\bibitem{Recycle2014} H. N. Phien, I. P. McCulloch, and G. Vidal,
Phys. Rev. B \textbf{91}, 115137 (2015).

\bibitem{PEPSfiniteT2015} Piotr Czarnik and Jacek Dziarmaga,
Phys. Rev. B \textbf{92}, 035152 (2015).

\bibitem{arXiv2015Orus} H. N. Phien, J. A. Bengua, H. D. Tuan, P. Corboz,
and R. Orus, arXiv:1503.05345.

\bibitem{TNMC2011} Ling Wang, Iztok Pizorn, Frank Verstraete,
Phys. Rev. B 83, 134421 (2011).

\bibitem{Biorthonormal2011} Yu-Kun Huang, Phys, Rev. E
\textbf{83}, 036702 (2011).

\bibitem{finiteIsing1949} Q. Kanfrnan, Phys. Rev. 76, 1232 (1949).

\bibitem{Kitaev2006} A. Kitaev, Ann. Phys. (NY) \textbf{321}, 2 (2006).

\bibitem{KHiPEPS2014} Juan Osorio Iregui, Philippe Corboz, and Matthias Troyer, Phys. Rev. B \textbf{90}, 195102 (2014).

\end{thebibliography}
\end{document}